# Interstellar Escape from Proxima b is Barely Possible with Chemical Rockets

A civilization in the habitable zone of a dwarf star might find it challenging to escape into interstellar space using chemical propulsion

———

By Abraham Loeb on April 8, 2018

Almost all space missions of our civilization were based so far on chemical propulsion. The fundamental limitation of chemical propulsion is easy to understand. The rocket is pushed forward by ejecting burnt fuel gases backwards through its exhaust. The characteristic composition and temperature of the burnt fuel set its exhaust speed to a typical value of a few kilometers per second. Momentum conservation implies that the terminal speed of the rocket is given by this exhaust speed times the natural logarithm of the ratio between the initial and final mass of the rocket. To exceed the exhaust speed by some large factor requires an initial fuel mass that exceeds the final payload mass by the exponential of this factor. Since the required fuel mass grows exponentially with terminal speed, it is not practical for chemical rockets to exceed a terminal speed that is more than an order of magnitude larger than the exhaust speed, namely a few tens of kilometers per second. Indeed, this was the speed limit of all spacecrafts launched so far by NASA or other space agencies.

By a fortunate coincidence, the escape speed from the surface of the Earth, 11 kilometers per second, and the escape speed from the location of the Earth around the Sun, 42 kilometers per second, are just around the speed limit attainable by chemical propulsion. This miracle allowed our civilization to design missions, such as *Voyager 1* and *2* or *New Horizons*, that will escape from the solar system into interstellar space. But is this fortune shared by other civilizations on habitable planets outside the solar system?

Life "as we know it" requires liquid water, which could exist on planets with a surface temperature and a mass similar to Earth. Surface heating is needed to avoid freezing of water into ice and an Earth-like gravity is needed to retain the planet's atmosphere, which is also essential since ice turns directly into gas in the absence of an external atmospheric pressure. The warning sign is just next door in the form of Mars, which has a tenth of an Earth mass and lost its atmosphere. Since the surface temperature of a warm planet is dictated by the flux of stellar irradiation, the distance of the habitable zone around any arbitrary star scales roughly as the square root of the star's luminosity. For the abundant population of low mass stars, the stellar luminosity scales roughly as the stellar mass to the third power. The escape speed scales as the square root of the stellar mass over the distance from the star. Altogether, these considerations imply that the escape speed from the habitable zone scales inversely with stellar mass to the power of one quarter. Paradoxically, the gravitational potential well is deeper in the habitable zone around lower mass stars. A civilization born near a dwarf star would need to launch rockets at a higher speed than we do in order to escape the gravitational pull of its star, even though the star is of lighter than the Sun.

As it turns out, the lowest mass stars happen to be the most abundant of them all. It is therefore not surprising that the nearest star to the Sun, Proxima Centauri, has 12% of the mass of the Sun. This star also hosts a planet, Proxima b, in its habitable zone at a distance that is 20 times smaller than the Earth-Sun separation. The escape speed from the location of Proxima b to interstellar space is about 65 kilometers per second. Launching a rocket from rest at that location requires the fuel-to-payload weight ratio to be larger than a few billions in order for the rocket to escape the gravitational pull of Proxima Centauri. In other words, freeing a speck weighing one gram of technological equipment from the position of Proxima b to interstellar space requires a chemical fuel tank that weighs millions of kilograms, similar to that used for liftoff of the Space Shuttle. Increasing the final payload weight to a kilogram, the scale of our smallest CubeSat, requires a thousand times more fuel than carried by the Space Shuttle.

This is bad news for technological civilizations in the habitable zone of dwarf stars. Their space missions would barely be capable of escaping into interstellar space using chemical propulsion alone. Of course, the extraterrestrials (ETs) can take advantage, as we do, of gravitational assists by optimally designing the spacecraft trajectory around their host star and surrounding planets. In particular, launching a rocket in the direction of motion of the planet would reduce the propulsion boost needed for interstellar escape down to the practical range of 30 kilometers per second. The ETs could also employ more advanced propulsion technologies, such as lightsails or nuclear engines.

Nevertheless, this global perspective should make us feel fortunate that we live in the habitable zone of a rare star as bright as the Sun. Not only that we have liquid water and a comfortable climate to maintain a good quality of life, but we also inhabit a platform from which we can escape at ease into interstellar space. We should take advantage of this fortune to find real estate on extrasolar planets in anticipation of a future time when life on our own planet will become impossible. This unfortunate fate will inevitably confront us in less than a billion years, when the Sun will heat up enough to boil all water off the face of the Earth. With proper planning we could relocate to a new home by then. Some of the most desirable destinations would be systems of multiple planets around low mass stars, such as the nearby dwarf star TRAPPIST-1 which weighs 9% of a solar mass and hosts seven Earth-size planets. Once we get to the habitable zone of TRAPPIST-1, there would be no rush to escape. Such stars burn hydrogen so slowly that they could keep us warm for ten trillion years, about a thousand times longer than the lifetime of the Sun.

# ABOUT THE AUTHOR

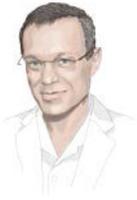

**Abraham Loeb**

Abraham Loeb is chair of the astronomy department at Harvard University, founding director of Harvard's Black Hole Initiative and director of the Institute for Theory and Computation at the Harvard-Smithsonian Center for Astrophysics. He serves as vice chair of the Board on Physics and Astronomy of the National Academies and chairs the advisory board for the Breakthrough Starshot project.

Credit: Nick Higgins